\documentclass[reprint,amsmath,amssymb,prl,longbibliography,
]{revtex4-2}
\usepackage{comment}
\usepackage{color}
\usepackage{graphicx}
\usepackage{subfigure}
\usepackage{dcolumn}
\usepackage{bm}
\usepackage{amsmath}
 
\usepackage{braket}
\usepackage{diagbox, eqparbox, hhline}
\usepackage{hyperref}
\DeclareMathAlphabet{\mathpzc}{OT1}{pzc}{m}{it}

\begin{document}

\title{Mesoscale phase separation of skyrmion-vortex matter \\
in chiral magnet-superconductor heterostructures
}

\author{Jos\'e F. Neto}%
\affiliation{%
Departamento de F\'isica, Universidade Federal de Pernambuco, Cidade Universit\'aria, 50670-901, Recife-PE, Brazil.
}%
\author{Cl\'ecio C. de Souza Silva}%
\email{clecio.cssilva@ufpe.br}
\affiliation{%
Departamento de F\'isica, Universidade Federal de Pernambuco, Cidade Universit\'aria, 50670-901, Recife-PE, Brazil.
}

\date{\today}

\begin{abstract}

We investigate theoretically the equilibrium configurations of many magnetic skyrmions interacting with many superconducting vortices in a superconductor-chiral magnet bilayer. We show that miscible mixtures of vortices and skyrmions in this system breaks down at a particular wave number for sufficiently strong coupling, giving place to remarkably diverse mesoscale patterns: gel, stripes, clusters, intercalated stripes and composite gel-cluster structures. We also demonstrate that, by appropriate choice of parameters, one can thermally tune between the homogeneous and density-modulated phases.


\end{abstract} 

\maketitle



The combination of superconductivity and magnetism in hybrid magnet-superconductor materials has lead to a remarkable range of novel phenomena and applications impossible to achieve when the constituent materials are taken isolatedly. 
Examples include dissipationless spin polarized currents~\cite{Yang2010,linder2015,eschrig2015}, spin valves with infinite magnetoresistance~\cite{Takahashi1999,Miao2008,Li2013}, a novel class of superconducting quantum devices~\cite{Yamashita2005,feofanov2010,Kapran2019}, and controlled manipulation of flux quanta (Abrikosov vortices) in superconductors~\cite{Lange2003,Milosevic2004,Clecio2007PRL,Vlasko-Vlasov2008,velez2008,Aladyshkin2009,Adamus2016,Vlasko-Vlasov2017,wang2018,lyu2020}.

Recently, there has been an increasing interest on the 
hybridization of topologically-protected magnetic textures called skyrmions and Abrikosov vortices in heterostructures comprising a superconducting (SC) film and a chiral magnetic (CM) layer~\cite{hals2016composite,dahir2019interaction,baumard2019,RaiJose2019,Rex2019,Palermo2020,Petrovic2021}. In these systems, a skyrmion and a nearby vortex interact with each other via their stray fields and/or via spin-orbit coupling (SOC) between the SC and CM layers. In the case of attractive interaction, they eventually form a bound pair with easily tunable dynamical properties
~\cite{hals2016composite,dahir2019interaction,baumard2019,RaiJose2019,Andriyakhina2021}. In particular, for strong SOC, a skyrmion-vortex pair can host localized Majorana bound states, which makes SC-CM hybrids a promising platform for future applications in topological quantum computing~\cite{Yang2016,Rex2019,Petrovic2021,mascot2021topological}. 
%
%
%

The physics of many vortices and many skyrmions in SC-CM hybrids is still poorly understood. The ability of skyrmions to influence the vortex dynamics in the superconducting layer has recently been corroborated experimentally~\cite{Palermo2020,Petrovic2021}.
However, the impact of the vortex system in the magnetic state of the CM layer remains unknown. As it is often the case in the physics of many interacting objects, the composite many-skyrmion-many-vortex matter can exhibit new emergent properties not found when the skyrmion crystal and/or the vortex lattice are treated individually. Therefore, it is necessary to take into account the feedback of both subsystems on each other so as to investigate possible collective effects resulting from their mutual interaction.


In this Letter, we investigate  equilibrium configurations of composite skyrmion-vortex matter in SC-CM heterostructures as a function of the main energy scales and characteristic lengths of the system. For weak skyrmion-vortex coupling, both vortices and skyrmions form homogeneous, quasi-triangular lattices as a result of their repulsive intraspecies interaction. 
We show that, upon increasing the skyrmion-vortex coupling energy, the homogeneous distribution of vortices and skyrmions becomes unstable with respect to density fluctuations of a particular wavelength irrespective of whether the skyrmion-vortex interaction is attractive or repulsive. This results in a series of density-modulated phases, such as clusters, stripes, and bubbles, similar to microphase separation phenomena observed in soft matter systems like block copolymers~\cite{bates2000block,Krishnamoorthy2006,Zhuang2016}, colloidal suspensions~\cite{stradner2004,osterman2007}, charged water-oil mixtures~\cite{Tasios2017}, and model systems with non-monotonic interactions~\cite{malescio2003,O-Reichhardt2010,Zhao2012,Varney2013,Komendova2013}. 




\begin{figure}[b]
\centering
\includegraphics[width=\linewidth]{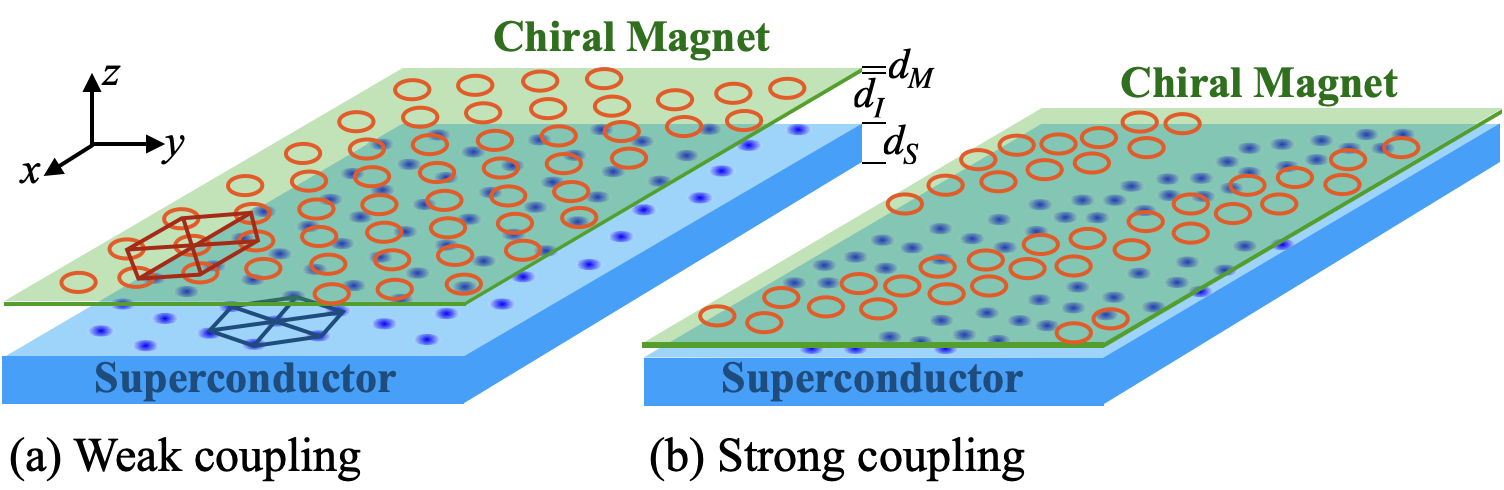}
\caption{Cartoon of the system geometry and illustration of possible distributions of skyrmions (rings) and vortices (dots) in two situations: (a) weak coupling between the SC and CM layers, 
where vortices and skyrmions form almost independent triangular lattices; (b) strong (repulsive) coupling, where the system self-organizes into a modulated phase.}
\label{fig.SystemCartoon}
\end{figure}
%


\emph{Model.}---We consider a thin, chiral ferromagnetic film or multilayer of thickness $d_\text{M}$ on top of a (SC) film of thickness $d_{\rm S}$, both exposed to an off-plane, external magnetic field $\bm{B}=B\hat{\bm{z}}$ and separated by an insulating layer of thickness $d_{\rm I}$ (see Fig.~\ref{fig.SystemCartoon}). We assume that $d_{\rm S}\ll\lambda$, where $\lambda$ is the London penetration depth of the superconductor, so that the number of vortices is essentially given by $N_\text{v}\simeq A_\text{film}B/\Phi_0$, where $A_\text{film}$ is the film surface area. In addition, the skyrmions in the CM layer are assumed to coexist with a ferromagnetic background and $d_\text{M}\ll d_{\rm S}$, so as to rule out the nucleation of vortices or vortex-antivortex pairs induced into the superconductor by the magnetic texture~\cite{baumard2019,dahir2019interaction}. The number $N_\text{s}$ of skyrmions in a CM layer can be controlled by a number of techniques, even for $B=0$~\cite{zhang2018direct,duong2019,meyer2019,brandao2019,guang2020creating}.  Therefore, to cover a wide range of possibilities, we will change $N_\text{s}$ and $N_\text{v}$ independently. We further assume that vortex-vortex and skyrmion-skyrmion distances are considerably larger than their respective core sizes and that $N_\text{s}$ and $N_\text{v}$ are conserved, which is a good approximation for vortices in the thin film limit and for skyrmions when the CM layer is prepared in the skyrmion crystal phase~\cite{SuppMat}.

With the considerations above
, vortices and skyrmions can be treated as particles and their total free energy can be expressed as ${\cal F}={\cal F}_0+{\cal F}_\text{int}$, where ${\cal F}_0$ is the self energy of all vortices and skyrmions and ${\cal F}_\text{int}$ is the superposition of all pair interactions, that is 
\begin{align}\label{eq.F}
{\cal F}_\text{int} = \frac{1}{2}\!\sum_{i,j=1}^{N_\text{v}}\!V_\text{vv}(r) + \frac{1}{2}\!\sum_{i,j=1}^{N_\text{s}}\!V_\text{ss}(r) 
+ \sum_{i=1}^{N_\text{v}}\sum_{j=1}^{N_\text{s}}\!V_\text{vs}(r) 
\end{align}
where $r=|\bm{r}_i-\bm{r}_j|$, with $\bm{r}_i$, $\bm{r}_j$ the positions of vortices or skyrmions in the $xy$ plane. 
The vortex-vortex interaction is modeled by Pearl's potential, which can be expressed as 
$ 
    V_\text{vv}(r) = (\epsilon_\text{vv}/2\pi)\int d^2\bm{k}\, {e^{i\bm{k}\cdot\bm{r}}}/(k^2+k/\Lambda),
$
where $\Lambda=2\lambda^2/d_\text{s}\gg\lambda$ is the Pearl length, $\epsilon_\text{vv}=\phi_0^2/\pi\mu_0\Lambda$, and $\phi_0=h/2e$ is the flux quantum~\cite{Brandt2009}. This is a long range repulsive potential, which for $r\ll\Lambda$ can be approximated by $\epsilon_\text{vv}\ln(\Lambda/r)$. For the skyrmion-skyrmion interaction  we use a modified Bessel function, which reflects the short range repulsive character of the potential for distances larger than the skyrmion radius $R_\text{sk}$~\cite{lin2013,foster2018,Brearton2020,Capic2020}: 
$V_\text{ss}(r) = \epsilon_\text{ss}K_0(r/\xi_\text{s})$,
where $\xi_\text{s}$ is the healing length of the spins outside the skyrmion core and  $\epsilon_\text{ss}$ has been recently estimated as $\epsilon_\text{ss}\sim 60Ad_\text{M}(R_\text{sk}/\xi_\text{s})^4$~\cite{Capic2020}, with $A$ the exchange stiffness.
Finally, we model the electromagnetic coupling between a vortex and a skyrmion by 
$V_\text{vs}(r) = {\epsilon_\text{vs}}/{(1+r^2/\lambda_\text{vs}^2)^2}$, where $\epsilon_\text{vs}$ can be either negative or positive, depending on whether the skyrmion is oriented parallel or antiparallel to the vortex, and $\lambda_\text{vs}\simeq0.8\lambda$ (see Ref.~\cite{RaiJose2019} and the Supplemental Material~\cite{SuppMat}). 
For $d_\text{I}=0$, this potential could also represent qualitatively the effect of proximity induced SOC between the SC and CM layers~\cite{hals2016composite,baumard2019}.


\emph{Mean-field theory.}---We analyze the stability of the homogeneous phase using a non-local mean-field approach where the free energy~\eqref{eq.F} is expressed as a functional of the densities coarse-grained over many inter-vortex and inter-skyrmion spacings~\cite{note_MFDFT}, 
$
\bar{\cal F}_\text{int}[n_\text{v},n_\text{s}] = \frac{1}{2}\sum_{\alpha,\beta}\int\! d^2\bm{r}\,d^2\bm{r}'\, n_\alpha(\bm{r}) n_\beta(\bm{r}') V_{\alpha\beta}(\bm{r}-\bm{r}'),
$
where $\alpha,\beta=\text{v, s}$, with the constraints $\int\!d^2\bm{r}\,n_\alpha(\bm{r})=N_\alpha$. The homogeneous distributions of vortices and skyrmions, $n_\text{v0}=N_\text{v}/A$ and $n_\text{s0}=N_\text{s}/A$, are a trivial solution of the minimization of this functional. 

\begin{figure*}[t]
\centering
\includegraphics[width=\linewidth]{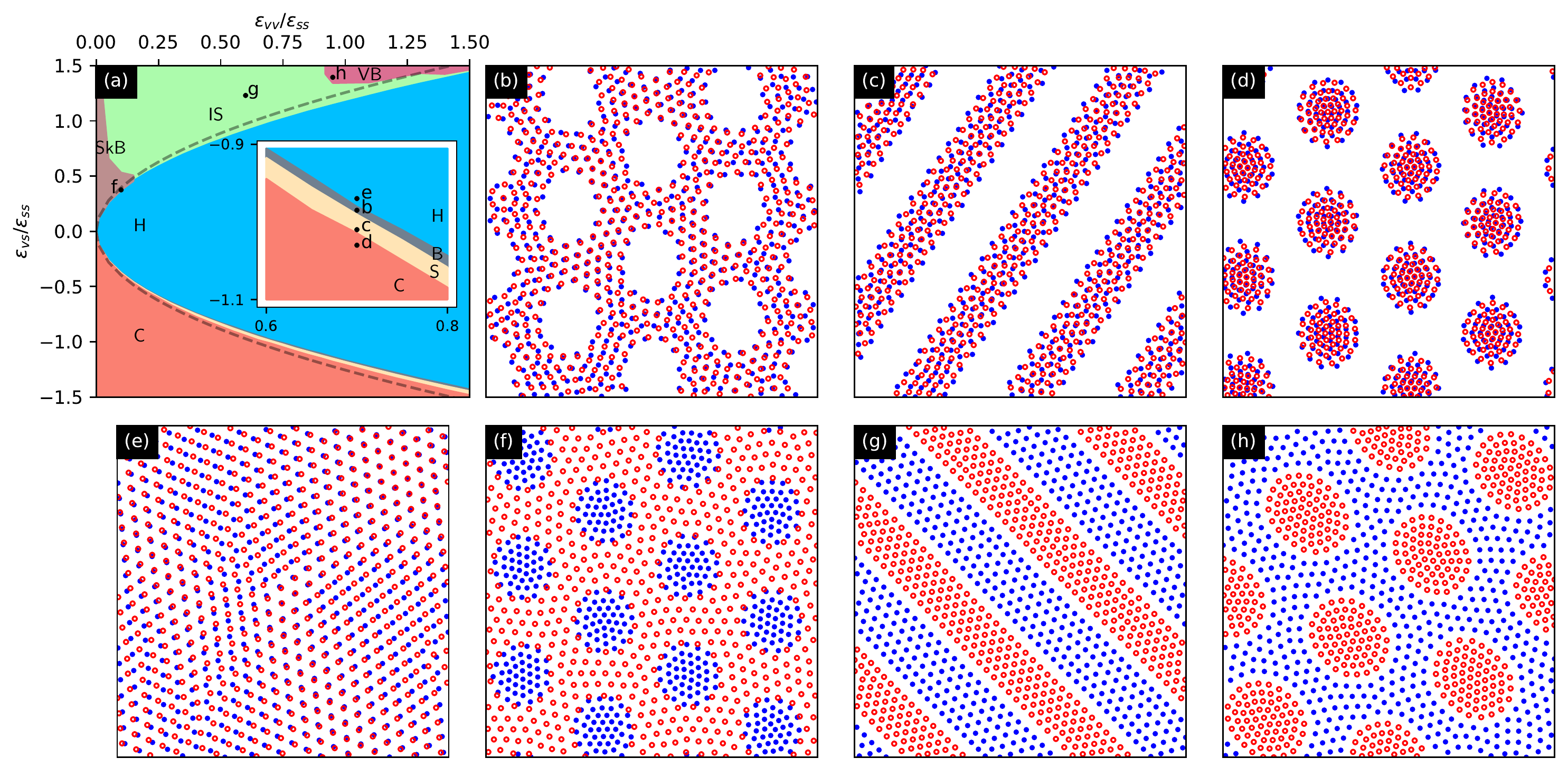}
\caption{Phase diagram (a) of the composite skyrmion-vortex system for $n_\text{v0}=3.55 \lambda_\text{vs}^{-2}$, $n_\text{s0}=3.77\lambda_\text{vs}^{-2}$, and $\xi_s=0.5\lambda_{vs}$, featuring the homogeneous phase (H) and six distinct microphases: bubbles (B), stripes (S), clusters (C), vortex-filled skyrmion bubbles (SkB), intercalated stripes (IS), and  skyrmion-filled vortex bubbles (VB). The dashed line is the theoretical stability boundary of the homogeneous phase calculated from Eqs.~\eqref{eq.inst_boundary} and \eqref{eq.CriticalWaveNumber2}. Inset: zoomed in phase diagram for better identification of phases B and S. (b)-(h) Representative configurations of vortices (blue dots) and skyrmions (red rings) for each phase as indicated in (a). }
\label{fig.Phases}
\end{figure*}

To evaluate the stability with respect to density fluctuations at specific wave vectors $\bm{k}$, 
we rewrite $\bar{\cal F}_\text{int}$ as a functional in Fourier space: 
\begin{equation}\label{eq.F_MF.Fourier}
\bar{\cal F}[\tilde{n}_\text{v},\tilde{n}_\text{s}] = {\cal F}_0 + \frac{1}{2}\sum_{\alpha,\beta}\int\! \frac{d^2\bm{k}}{4\pi^2}
\tilde{n}_\alpha(-\bm{k})\tilde{n}_\beta(\bm{k})\tilde{V}_{\alpha\beta}({k}),
\end{equation}
where $\tilde{f}(\bm{k})=\int d^2\bm{r}\, f(\bm{r})e^{-i\bm{k}\cdot\bm{r}}$. Stability requires that the Hessian matrix 
$
{\cal D}_{\alpha\beta}(\bm{k}) = \frac{\delta^2 \bar{\cal F}_\text{int}}{\delta\tilde{n}_\alpha(-\bm{k})\delta\tilde{n}_\beta(\bm{k})} = \tilde{V}_{\alpha\beta}({k})
$
be positive definite, i.e. $\det {\cal D}_{\alpha\beta}(\bm{k})=\tilde{V}_\text{vv}({k})\tilde{V}_\text{ss}({k})-\tilde{V}_\text{vs}({k})^2>0$, for all $\bm{k}$. However, this condition can be violated at a single non-zero wave number, ${k}_*$, satisfying the conditions
\begin{equation}
\label{eq.InstabCond}
\left[\det{\cal D}_{\alpha\beta}({k})\right]_* = 0 ,\quad\left[\frac{d}{d{k}}\det{\cal D}_{\alpha\beta}({k})\right]_* = 0,
\end{equation}
where the subscript $*$ means evaluation at ${k}_*$. The first equation establishes a simple relation between the energy scales at the stability boundary,
\begin{equation}
\label{eq.inst_boundary}
|\epsilon_\text{vs}| = \gamma_*\sqrt{\epsilon_\text{vv}\epsilon_\text{ss}},
\end{equation}
where $\gamma_*=\sqrt{\tilde{v}_\text{vv}(k_*)\tilde{v}_\text{ss}(k_*)}/\tilde{v}_\text{vs}(k_*)$ is a dimensionless constant, with $\tilde{v}_{\alpha\beta}\equiv\tilde{V}_{\alpha\beta}/\epsilon_{\alpha\beta}$. In other words, the homogeneous phase 
breaks down when $|\epsilon_\text{vs}|> \gamma_*\sqrt{\epsilon_\text{vv}\epsilon_\text{ss}}$, irrespective of whether the vortex-skyrmion interaction is attractive or repulsive, as long as a non-zero solution for $k_*$ exists. 
The combination of both conditions in \eqref{eq.InstabCond} leads to an implicit equation relating  $k_*$ to all three length scales,
\begin{equation}
\label{eq.CriticalWaveNumber1}
\left[
\frac{\tilde{v}_\text{vv}^\prime}{\tilde{v}_\text{vv}} + 
\frac{\tilde{v}_\text{ss}^\prime}{\tilde{v}_\text{ss}} - 
2\frac{\tilde{v}_\text{vs}^\prime}{\tilde{v}_\text{vs}}
\right]_* = 0,
\end{equation}
where the prime means derivative with respect to $k$. This general result can be applied to other models of the interaction potentials. For the specific model considered here, $\tilde{v}_\text{vv} = {2\pi}/(k^2+k/\Lambda)$, $\tilde{v}_\text{ss} = 2\pi/(k^2+\xi_\text{s}^{-2})$, and $\tilde{v}_\text{vs} = \pi\lambda^3_\text{vs}kK_1(k\lambda_\text{vs})$, 
and Eq.~\eqref{eq.CriticalWaveNumber1} reads
\begin{equation}
\label{eq.CriticalWaveNumber2}
\frac{2k_*+\Lambda^{-1}}{k_*(k_*+\Lambda^{-1})}
+ \frac{2k_*}{k_*^2+\xi_\text{s}^{-2}} = 
2\lambda_\text{vs}\frac{K_0(k_*\lambda_\text{vs})}{K_1(k_*\lambda_\text{vs})}.
\end{equation}
Note that the left-hand side of this equation decreases monotonically toward zero, while the right-hand side increases monotonically toward $2\lambda_\text{vs}$. Therefore, no matter the combination of values of the length scales $\Lambda$, $\xi_\text{s}$ and $\lambda_\text{vs}$, there is always a single nonzero solution for $k_*$. Upon fulfillment of condition~\eqref{eq.inst_boundary}, the homogeneous state becomes unstable with respect to density fluctuations of wavelength $2\pi/k_*$. 




\emph{Numerical simulations.}---To investigate further the structure of the modulated phases and analyze the phase boundaries beyond mean-field approximation, we introduce coupled Langevin equations of motion for vortices and skyrmions:
\begin{eqnarray}\label{eq.Langevin_v}
     \eta_\text{v}\frac{d\bm{r}_{vi}}{dt} &=& -\frac{\delta{\cal F}_\text{int}}{\delta\bm{r}_{vi}}+\bm{\gamma}_{\text{v}i}(t), \\
     \label{eq.Langevin_s}
     (\eta_\text{s}+\bm{G}\times)\frac{d\bm{r}_{si}}{dt} &=& -\frac{\delta{\cal F}_\text{int}}{\delta\bm{r}_{si}}+\bm{\gamma}_{\text{s}i}(t), 
\end{eqnarray}
where $\eta_\text{v}$ ($\eta_\text{s}$) is the vortex (skyrmion) viscous drag coefficient and $\bm{G}=G\hat{\bm{z}}$ is the gyromagnetic vector~\cite{RaiJose2019}. The Gaussian noises $\bm{\gamma}_{\text{v}i}(t)$ and $\bm{\gamma}_{\text{s}i}(t)$ satisfy 
$\langle\gamma^\mu_{\alpha i}(t)\rangle=0$ and $\langle\gamma^\mu_{\alpha i}(t)\gamma^\nu_{\alpha j}(t')\rangle=\Gamma_\alpha\delta_{\mu\nu}\delta_{ij}\delta(t-t')$, where $\mu,\nu=x,y$ and $\alpha=\text{v},\text{s}$. In thermal equilibrium, $\Gamma_\alpha=2\eta_\alpha k_BT$~\cite{trugman1982,lin2013}. For the purpose of minimizing the full free energy~\eqref{eq.F}, we set arbitrarily $\eta_\text{v}=\eta_\text{s}$ and $G=0$, and integrate Eqs.~\eqref{eq.Langevin_v} and \eqref{eq.Langevin_s} on a square simulation box of side $L$ with periodic boundary conditions while slowly reducing the noise amplitude. Further simulation details and a discussion about the effect of $\eta_\text{v}$, $\eta_\text{s}$, and $G$ on the relaxation toward equilibrium are given in the Supplemental Material~\cite{SuppMat}.

For a general view of the possible phases, we 
minimize the free energy~\eqref{eq.F} for $N_\text{v}=512$ and $N_\text{s}=544$,  $L=12\lambda_\text{vs}=9.6\lambda$, and $\xi_s=0.5\lambda_{vs}$, thoroughly exploring the parameter space defined by $\epsilon_\text{vv}$ and $\epsilon_\text{vs}$, treated as independent variables. The resulting phase diagram is shown in Fig.~\ref{fig.Phases}. Notice that the boundary of the homogeneous (H) phase 
lies close to the theoretical instability line, Eq.~\eqref{eq.inst_boundary}, and is perfectly symmetric with respect to the sign of $\epsilon_\text{vs}$. 
Within this boundary (small $|\epsilon_\text{vs}|$), both vortices and skyrmions form homogeneous and slightly distorted triangular lattices~\cite{SuppMat}. 
For larger $|\epsilon_\text{vs}|$, the homogeneous phase gives place to  surprisingly diverse microphases. For attractive skyrmion-vortex potential ($\epsilon_\text{vs}<0$), we observed a narrow region of vortex-skyrmion bubble lattices (B) in the vicinity of the H-phase boundary, followed by commensurate stripes (S), and then commensurate cluster lattices (C), which ultimately dominate the large $|\epsilon_\text{vs}|$ region of the phase diagram. 
In contrast, for repulsive skyrmion-vortex potential ($\epsilon_\text{vs}>0$), the switching between the different microphases is strongly dependent on the ratio $\epsilon_\text{vv}/\epsilon_\text{ss}$. For large (small) $\epsilon_\text{vv}/\epsilon_\text{ss}$, the vortex (skyrmion) lattice becomes stiffer forcing skyrmions (vortices) to form compact clusters within vortex (skyrmion) bubbles [VB (SkB)]. For intermediate $\epsilon_\text{vv}/\epsilon_\text{ss}$, vortices and skyrmions come to terms and form intercalated stripes (IS). 

Going beyond the validity limit of the mean-field approximation, we performed a series of simulations for other skyrmion densities, down to $n_\text{s0}=0.89\lambda_\text{vs}^{-2}=0.22\xi_\text{s}^{-2}$. The results, presented in the Supplemental Material~\cite{SuppMat}, demonstrate that the microphases occupy a larger region of the phase diagram for lower $n_\text{s0}$, so that Eq.~\eqref{eq.CondTscan} can be viewed as an upper-limit estimate of the breakdown of the homogeneous phase.



%
\begin{figure}[t]
\centering
\includegraphics[width=0.9\linewidth]{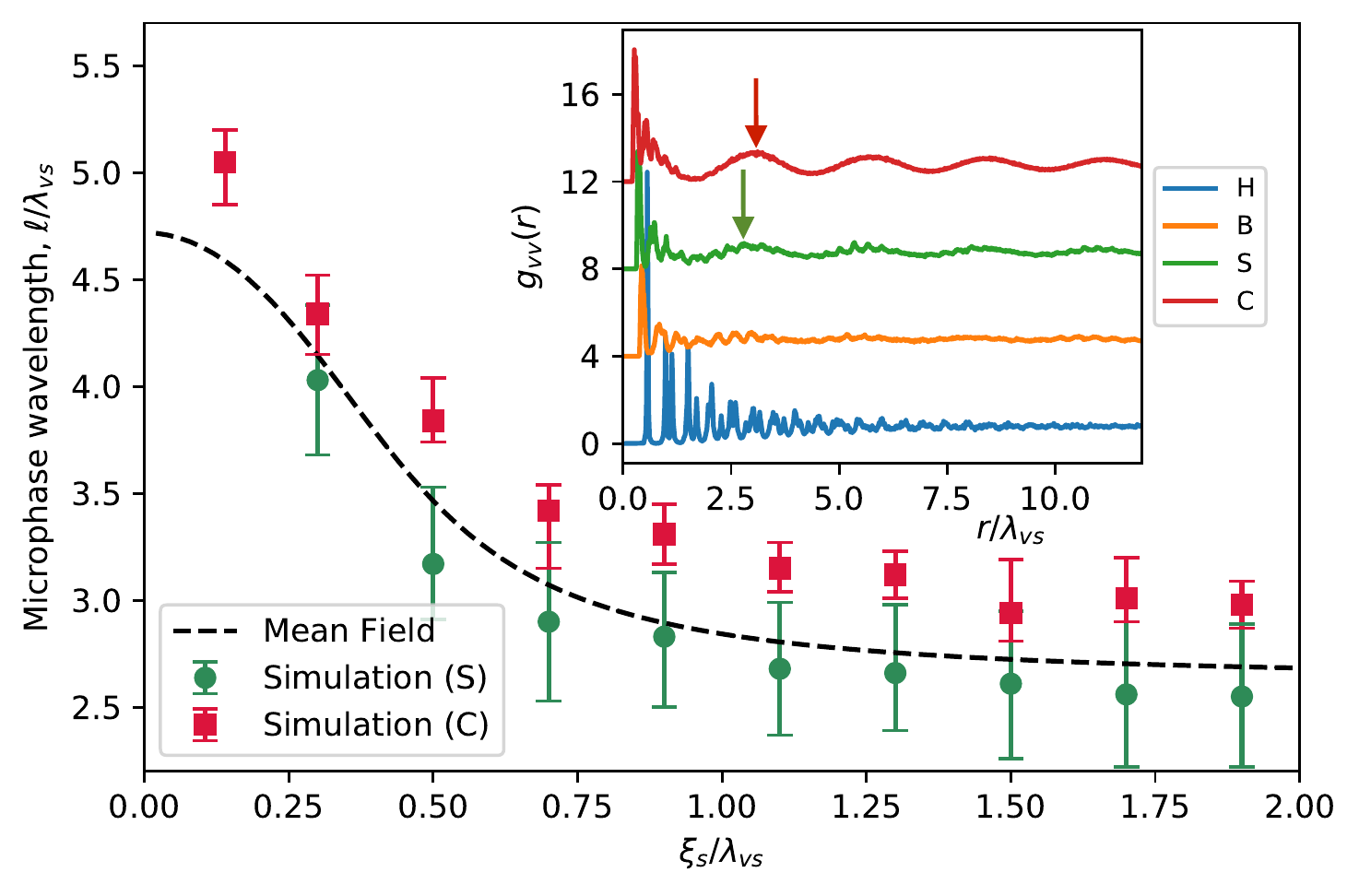}
\caption{Wavelength of the stripe and cluster phases simulated for different $\xi_s$. Here, $L=24\lambda_\text{vs}$,  $N_\text{v}=N_\text{s}=2048$, $\epsilon_\text{vs}=\epsilon_\text{ss}$ and $\epsilon_\text{vv}$ is chosen as to keep close to the instability line. The full line corresponds to the critical wavelength, $2\pi/k_*$, calculated from Eq.~\eqref{eq.CriticalWaveNumber2}. Inset: radial vortex-vortex correlation function $g_\text{vv}(r)$ of configurations representative of the H, B, S, and C phases. The arrows indicate the microphase wavelength of the respective configuration. 
}
\label{fig.TemperatureDependence}
\end{figure}

The structure of the microphases is revealed as smooth mesoscale oscillations in the radial correlation function of skyrmions and vortices, $g_{\alpha\alpha}(r)$, a quantity readily accessible in e.g.\ neutron diffraction experiments. Here, we use $g_\text{vv}(r)$ for an accurate estimate of the typical wavelength of the microphases in the vicinity of the instability line as a function of $\xi_s$ (see Fig.~\ref{fig.TemperatureDependence}). 
To minimize finite size effects, we performed simulations on a large simulation box of size $L=24\lambda_\text{vs}$ for $N_\text{v} = N_\text{s}=2048$. For each $\xi_\text{s}$, we fixed $\epsilon_\text{vs}=1.0\epsilon_\text{ss}$ and chose $\epsilon_\text{vv}$ in a way as to stay close to the boundary of the homogeneous phase. As seen in Fig.~\ref{fig.TemperatureDependence}, the numerical data lies remarkably close to the critical wavelength $\ell_*=2\pi/k_*$, suggesting that the structure of the microphases is determined precisely by the wavenumber of the density fluctuations responsible for the breakdown of the homogeneous phase.

\emph{Experimental accessibility.}---To evaluate the experimental accessibility of the vortex-skyrmion microphases it is necessary to determine the temperature dependence of the several quantities involved. To be specific, we consider systems comprising a low-$T_c$ superconducting film and a chiral magnetic multilayer with Curie temperature $T_C\gg T_c$. In this case, the operating temperature satisfies $T<T_c\ll T_C$, in which case 
$\xi_\text{s}$ and $\epsilon_\text{ss}$ are essentially constant~\cite{Tomasello2018}. In contrast, $\lambda$, $\epsilon_\text{vv}$, and $\epsilon_\text{vs}$ are strongly temperature-dependent. In particular, close to the superconducting critical temperature, $T_c$, $\epsilon_\text{vv}, \epsilon_\text{vs}\sim 1/\lambda^2(T)$ and thereby $\gamma^2=\epsilon_\text{vs}^2/(\epsilon_\text{vv}\epsilon_\text{ss})\sim1/\lambda^2(T)$ (see Supplemental Material~\cite{SuppMat}). In this limit, one also has $\Lambda(T)\gg\lambda_\text{vs}(T)\gg\xi_\text{s}$ and Eq.~\eqref{eq.CriticalWaveNumber2} can be considerably simplified leading to  $\gamma_*^2\simeq15.76\,\xi_\text{s}^2/\lambda^2(T)$. Therefore, for $T$ close to $T_c$, $\gamma$ and $\gamma_*$ have the same temperature dependence and the condition for vortex-skyrmion microphase reduces to $\gamma_*^2(0)<\gamma^2(0)$ or:
\begin{equation}\label{eq.CondTscan}
    15.76\frac{\xi^2_\text{s}}{\lambda^2(0)} < \frac{\epsilon^2_\text{vs}(0)}{\epsilon_\text{vv}(0)\epsilon_\text{ss}} \sim \frac{\pi}{15} \frac{\xi_\text{s}^4(d_\text{S}{\cal R}_1 + R_\text{sk}{\cal R}_2)^2}{\ell_{\rm ex}^2\lambda^2(0)d_\text{M}d_\text{S}},
\end{equation} 
The right-hand-side is a rough estimate of $\gamma^2(0)$ for small skyrmion radius, ${\cal R}_1$ and ${\cal R}_2$ are dimensionless geometric factors, and 
$\ell_{\rm ex}=\sqrt{2A/\mu_0M^2_\text{s}}$ is the exchange length of the CM layer, with $M_\text{s}$ its saturation magnetization~\cite{SuppMat}. Away from the $T\rightarrow T_c$ limit, the scaling $\gamma, \gamma_* \sim \lambda^{-2}$ is no longer valid and the temperature dependence of each of these quantities has to be determined numerically~\cite{SuppMat}. 

\begin{figure}[t]
\centering
\includegraphics[width=\linewidth]{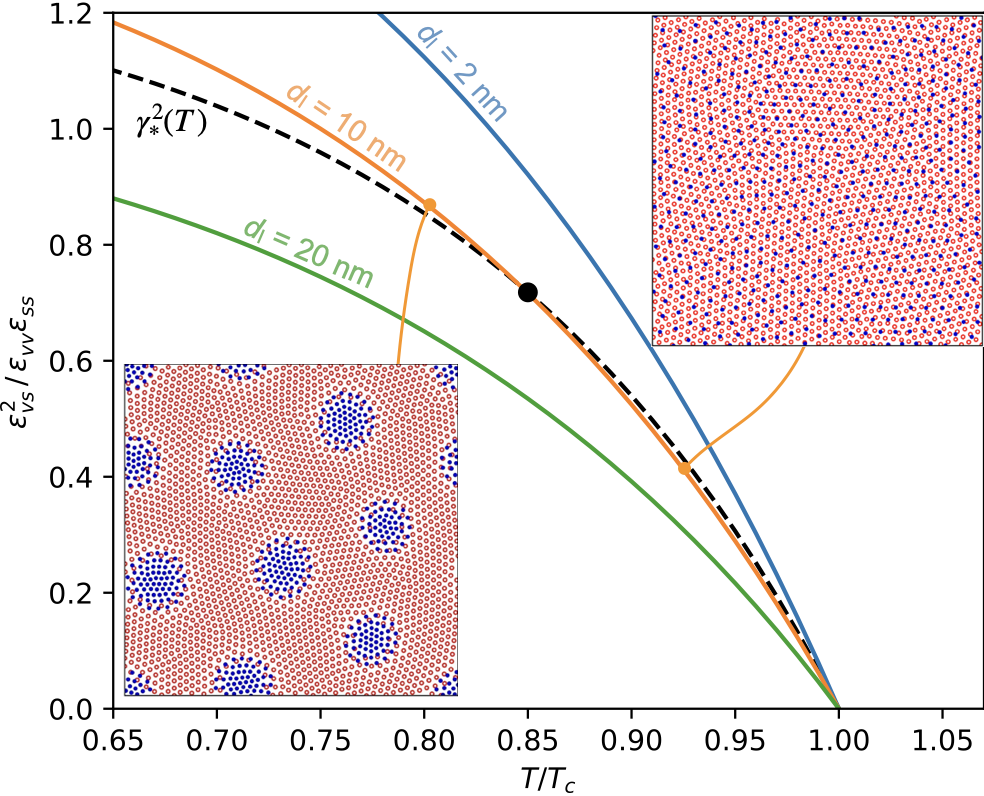}
\caption{Temperature dependence of  $\gamma^2=\epsilon_\text{vs}^2/\epsilon_\text{vv}\epsilon_\text{ss}$ calculated for a superconductor-insulator-ferromagnet trilayer considering three different values of $d_\text{I}$. Also shown is the instability line $\gamma_*^2(T)$ (dashes). When $\gamma>\gamma_*$ the homogeneous phase decays to a microphase. The insets are field-cooled vortex-skyrmion configurations for $d_\text{I}=10$ nm at $T_1=0.80T_c$ and $T_2=0.925T_c$.
}
\label{fig.TemperatureDependence}
\end{figure}
%


For a concrete example, we consider three CM-SC heterostructures identical to each other except for the thickness of their insulating layer: $d_\text{I}=2$, 10, and 20 nm. %
For the SC layer, we take $d_\text{S}=20$ nm and  $\lambda(T)=\lambda(0)/\sqrt{1-(T/T_c)^4}$, with $\lambda(0)=150$ nm. For the CM layer we choose $d_\text{M}=10$ nm,  $R_\text{sk}=40$ nm, $\xi_\text{s}=50$ nm, $A=10.45$ pJ/m and $M_\text{s}=1.0$ MA/m, typical of chiral magnetic multilayers~\cite{Palermo2020,Petrovic2021}.  This amounts to $\epsilon_\text{vv}(0)=0.482$ aJ and $\epsilon_\text{ss}=2.57$ aJ. The vortex-skyrmion coupling constant $\epsilon_\text{vs}$ was computed as a function of temperature for each $d_I$~\cite{SuppMat}. Notice that the estimated threshold magnetization for skyrmion-induced vortex-antivortex pairs in the SC layer is~\cite{dahir2019interaction}  $M_\text{th}=\phi_0\ln(\Lambda/2\xi)/(0.86\pi^2\mu_0d_\text{M}R_\text{sk})=2.63$ MA/m $>M_s$, for $T=0$, $d_I=0$, and assuming $\xi(0)=5$ nm for the superconducting coherence length. Hence vortex-antivortex pairs induced by the skyrmions can be ruled out in this example. In Fig.~\ref{fig.TemperatureDependence}, we plot the ratio $\gamma^2=\epsilon_\text{vs}^2/(\epsilon_\text{vv}\epsilon_\text{ss})$ as a function of reduced temperature $T/T_c$ for all three samples and the instability line $\gamma^2_*(T)$. For small $d_I$, the homogeneous phase is unstable as soon as the SC film enters the superconducting state, while for large $d_I$, the homogeneous phase is stable in the full temperature range of the superconducting state. Interestingly, for the intermediate case, $d_\text{I}=10$ nm, $\gamma(T)$ crosses the critical line at $T_*=0.85T_c$: the system is homogeneous at high temperatures, but breaks down into a modulated phase at $T<T_*$
We further support these results with numerical simulations of $N_\text{v}=512$ and $N_\text{s}=2048$ on a square simulation box of size $L=2.0~\mu$m, thus corresponding to an external magnetic field $B=265$ mT. For all three values of $d_\text{I}$ we simulate field cooling processes from the normal state down to the target temperatures $T_1=0.80T_c$ and $T_2=0.925T_c$ by first initializing the skyrmions as a triangular lattice (previously annealed in the absence of vortices) and placing the vortices at random positions. From this point on, the binary system is allowed to relax toward a stationary state at the target temperature. The stationary vortex-skyrmion configurations obtained were found to be modulated at both $T_1$ and $T_2$ for $d_I=2$ nm, modulated at $T_1$ and homogeneous at $T_2$ for $d_I=10$ nm, and homogeneous at both temperatures for $d_I=20$ nm, thus in excellent agreement with the mean-field prediction.



\emph{Summary.}---In conclusion, we have demonstrated that a binary magnetic skyrmion-superconducting vortex system can exhibit emergent density-modulated phases when the coupling between the superconductor and the ferromagnet is strong enough. For moderate to high skyrmion and vortex densities, the numerical simulations are in excellent agreement with the analytical results for the instability boundary of the homogeneous phase and  wavelength of the density modulations, calculated from mean-field density functional theory. We also derived an approximate inequality which allows one to tune material and geometrical parameters for the observation of vortex-skyrmion mesophases. In particular, for intermediate separations between the superconductor and the chiral magnet, it is possible to tune between homogeneous and microphase vortex-skyrmion configurations by changing the temperature of the system. These findings shed light on the general problem of pattern formation in binary mixtures and provide new insights on composite vortex-skyrmion systems which can be useful for future spintronics and quantum computing applications.


We are grateful to M.M. Milosevic and R.M. Menezes for useful discussions. This work was supported by the Brazilian Agencies FACEPE, under Grant No. APQ-0198-1.05/14, CAPES, and CNPq.

\bibliography{references}

\begin{thebibliography}{54}
\bibitem [{not()}]{note_MFDFT}%
  \BibitemOpen
  \href@noop {} {}\bibinfo {note} {This approach is expected to be accurate
  when the mean spacing between particles of the same species is smaller than
  the relevant characteristic lengths of the interactions, that is, when
  $a_\text{v}\equiv n_\text{v}^{-1/2}\lesssim\Lambda,\lambda_\text{vs}$ and
  $a_\text{s}\equiv
  n_\text{s}^{-1/2}\lesssim\xi_\text{s},\lambda_\text{vs}$}\BibitemShut
  {NoStop}%
\end{thebibliography}

\end{document}


\title{Supplemental Material: Mesoscale phase separation of skyrmion-vortex \\
matter in chiral magnet-superconductor heterostructures
}

\author{Jos\'e F. Neto}%
\author{Cl\'ecio C. de Souza Silva}%
\affiliation{%
Departamento de F\'isica, Universidade Federal de Pernambuco, Cidade Universit\'aria, 50670-901, Recife-PE, Brazil.
}%

\date{\today}

\maketitle

\section{Details of the minimization procedure}
\label{sec.Simul}

For a given set of parameters, we minimize the free energy~(1) following a simulated annealing scheme where we initialize all vortices and skyrmions at random positions and adjust the noise amplitude $\Gamma$ in Eqs.~(7) and (8) as to guarantee  both skyrmions and vortices are in a molten state. Then, we solve Eqs.~(7) and (8) using the Euler-Maruyama  method while slowly reducing $\Gamma$  to zero.  The typical number of time steps for a complete annealing procedure at a given point of the parameter space is $5\times10^5$. To accelerate the determination of the phase boundaries, we executed this simulated annealing procedure first in a sparse set of points in the parameter plane. Once two distinct phases at nearby points are detected, we perform a new annealing at the midpoint and identify the phase there. This bisection scheme is then repeated iteratively (typically 4 to 5 times) until convergence to an accuracy of three decimal places is reached.

\begin{figure*}[b]
\centering
\includegraphics[width=0.9\linewidth]{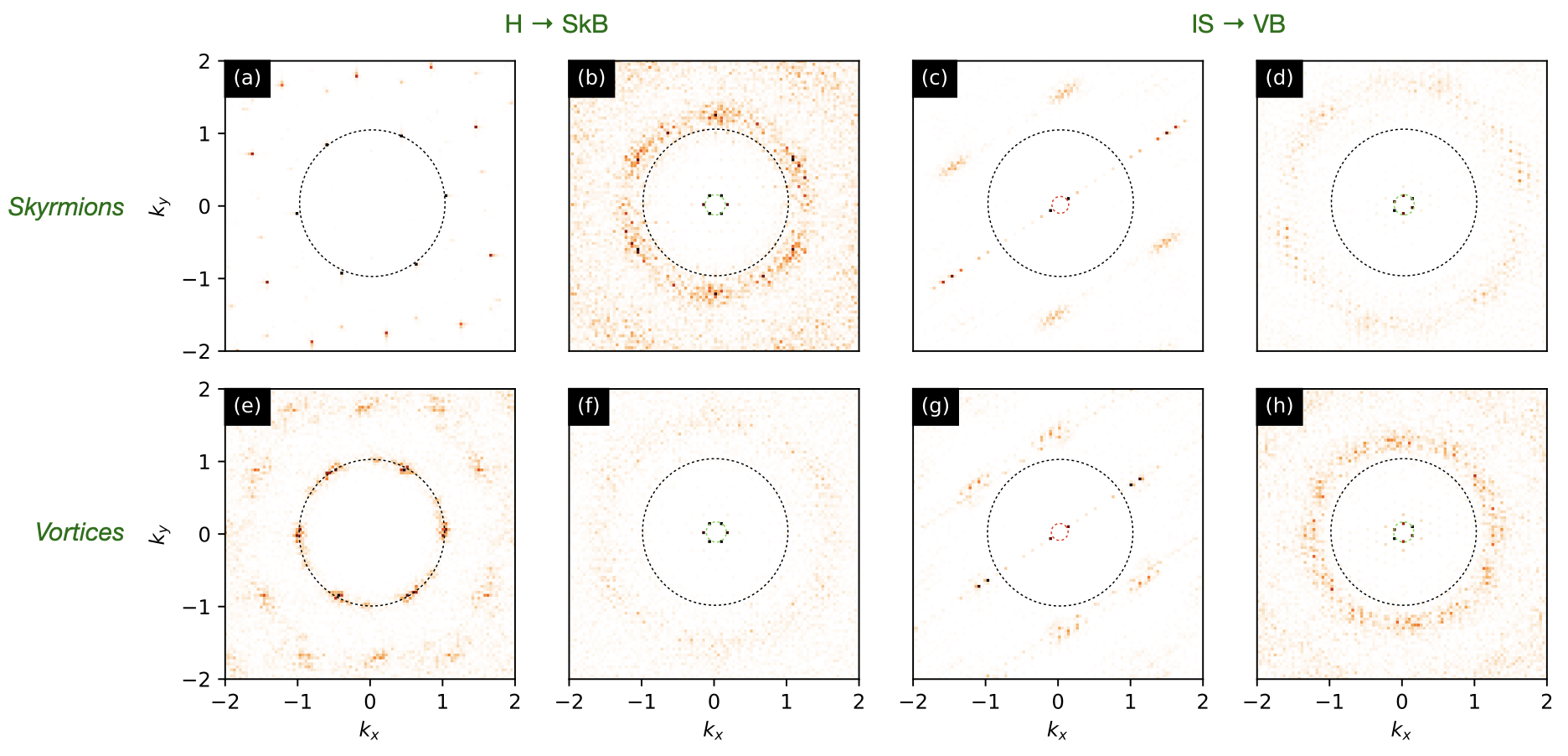}
\caption{Density plots of the structure factor $S(\bm k)$ for skyrmions (a)-(d) and vortices (e)-(h) at four points of the phase diagram of Fig. 2: (a, e) $\epsilon_{vv} = 0.15\epsilon_{ss}$ and $\epsilon_{vs} = 0.456\epsilon_{ss}$ (phase H); (b, f) $\epsilon_{vv}=0.15$ and $\epsilon_{vs} = 0.457\epsilon_{ss}$ (phase SkB); (c, g) $\epsilon_{vv} = 1.100\epsilon_{ss}$ and $\epsilon_{vs} = 1.348\epsilon_{ss}$ (phase IS); (d, h) $\epsilon_{vv}=1.100$ and $\epsilon_{vs} = 1.349\epsilon_{ss}$ (phase VB). The dotted circles are guides to the eye. Their radii indicate selected reciprocal lattice constants $k_1=\frac{4\pi}{\sqrt{3}a_v}$ (black), where $a_v$ is the theoretical lattice constant of a perfect vortex lattice, $k_2=\frac{4\pi}{\sqrt{3}\ell}$ (green), and $k_3=\frac{2\pi}{\ell}$ (red), where $\ell$ is the main periodicity of the corresponding mesoscale phase. $k_x$ and $k_y$ are in units of $k_1$. 
}
\label{fig.S(k)}
\end{figure*}

All phases, including those at the close vicinity of the phase boundaries, could be identified unambiguously by visual inspection of the configurations. In situations where the system is subjected to disorder, like in actual experiments, the structure factor $S(\bm{k})$~\cite{chaikin2000principles} can be used as an order parameter to better identify the different types of correlations present in each phase. In Fig.~\ref{fig.S(k)}, we show density plots of the structure factor of vortices and skyrmions for some selected points of the phase diagram shown in Fig. 2. Notice that the points in phases H (a,e) and SkB (b,f) were chosen very close to each other. Even then, the features of each phase can be clearly distinguished. The same applies to the points in phases IS (c,g) and VB (e,h). For instance, across the H$\rightarrow$SkB transition the Bragg peaks corresponding to the homogeneous triangular lattice are essentially lost and new Bragg peaks appear at reciprocal lattice vectors not of the individual particle arrangements but the mesoscale density modulation, which has considerably larger lattice constant. These Bragg peaks look the same for skyrmions, which form bubbles in the SkB phase, and vortices, which form clusters. The ring visible in the $S(\bm{k})$ data of the skyrmion bubble phase is a distinctive feature, not clearly present in the cluster phase. It is indicative that the skyrmions are in a slightly compressed liquid-like state. Similar behavior is observed in the VB phase, but with the roles of vortices and skyrmions interchanged. Finally, the stripe structure of vortices and skyrmions in the IS phase are clearly identified in the $S(\bm{k})$ data as the twofold Bragg spots appearing at small values of the reciprocal lattice constants.

The $S(\bm{k})$ data also reveal that the vortex and skyrmion lattices in the homogeneous phase can present different degrees of order. In the example of panels (a) and (e), skyrmons form a well ordered triangular lattice because, for this set of parameters, they interact more strongly than vortices, which, being softer, have to adapt their own correlations to those of the skyrmions, resulting in a more distorted lattice.

\section{Skyrmion-vortex coupling energy}

We consider heterostructures where the SC and CM layers are electrically isolated from each other by an insulating layer. In this case, the vortex-skyrmion interaction energy is given solely by the interaction between their stray fields and can be approximated by $V_\text{vs}(r_\text{vs})=\epsilon_\text{vs}/(1+r_\text{vs}^2/\lambda_\text{vs}^2)^c$, where $\lambda_\text{vs}\simeq0.8\lambda$ and $c\simeq2$~\cite{RaiJose2019}. To estimate $\epsilon_\text{vs}$ and determine its temperature dependence, we assume a rigid skyrmion profile. In this case
\begin{equation}\label{eq.evs}
\epsilon_{vs} = \mathcal{F}_\text{Zeeman}(r_\text{vs}=0) - \mathcal{F}_\text{Zeeman}(r_\text{vs}=\infty),
\end{equation}
where $\mathcal{F}_\text{Zeeman} = -\int \bm{B}_\text{v}\cdot d\bm{m} = -M_s\int\bm{B}_\text{v}\cdot\hat{\bm{n}}\,d^2\bm{r}\,dz$ is the Zeeman contribution to the free energy of SC-CM system containing one vortex and one skyrmion. Here $\bm{r}$ is the position of the spin in the $xy$ plane, $z$ is the vertical distance of the spin with respect to the upper surface of the SC layer, $\hat{\bm{n}}$ is the spin orientation vector at $(\bm{r},z)$, and $\bm{B}_\text{v}$ is the stray field induced by the vortex at $(\bm{r},z)$.  Notice that the integration in $z$ is to be taken across the CM thickness, that is, from $d_\text{I}$ to $d_\text{I}+d_\text{M}$.

We consider a right-handed N\'eel skyrmion with spin profile  $\hat{\bm{n}}=\hat{z}\cos\theta(\rho) - \hat{r}\sin\theta(\rho)$, where $\theta$ is the spin orientation angle with respect to the magnetic moment of the background ferromagnetic state and $\rho$ is the position of the spin with respect to the skyrmion center. $\theta(\rho)$ can be well approximated by the following ansatz~\cite{Romming2015}:
\begin{align}\label{eq.theta}
 \theta = \pi - \arcsin\left[\tanh\left(\frac{\rho - R_\text{sk}}{\xi_s}\right)\right]
 - \arcsin\left[\tanh\left(\frac{\rho + R_\text{sk}}{\xi_s}\right)\right].
\end{align}
This formula describes a spin profile where $\theta$ goes from $\pi$ at the skyrmion center to $\pi/2$ within a distance $\sim R_\text{sk}$ and then approaches exponentially the ferromagnetic state ($\theta=0$) with a characteristic length $\xi_s$. 

The radial and $z$ components of $\bm{B}_\text{v}$ evaluated outside the superconducting region ($z>0$) are given by~\cite{carneiro2000vortex} 
\begin{subequations}\label{Bz}
\begin{align}
    \label{eq0a}
    B_r(r,z)=&\frac{\phi_0}{2\pi\lambda^2}\int_0^{\infty}dk\frac{kJ_1(kr)}{k^2+\lambda^{-2}}f(k,z), \\ 
    \label{eq0b}
    B_z(r,z)=&\frac{\phi_0}{2\pi\lambda^2}\int_0^{\infty}dk\frac{kJ_0(kr)}{k^2+\lambda^{-2}}f(k,z), 
\end{align}
\end{subequations}
where $J_\nu(x)$ is the $\nu$-th order Bessel function of the first kind and
$$
f(k,z)=\tau e^{-kz}\frac{(k+\tau)e^{\tau d_\text{S}}+(k-\tau)e^{-\tau d_\text{S}}-2k}{(k+\tau)^2e^{\tau d_\text{S}}-(k-\tau)^2e^{-\tau d_\text{S}}}, \quad \text{with} \quad \tau=\sqrt{k^2+\lambda^{-2}}.
$$ 
The coupling energy $\epsilon_\text{vs}$ is then numerically calculated upon substitution of Eqs.~\eqref{eq.theta}, \eqref{eq0a}, and \eqref{eq0b} in Eq.~\eqref{eq.evs} for different temperatures following the two-fluid model result: $\lambda(T)=\lambda(0)/\sqrt{1-(T/T_c)^4}$. The curves of $\gamma^2=\epsilon_\text{vs}^2/(\epsilon_\text{vv}\epsilon_\text{ss})$ as a function of temperature depicted in Fig. 4 were calculated following this procedure.

\subsection{Estimate of $\epsilon_\text{vs}$ in the limit $T\rightarrow T_c$}

The complexity of  Eq.~\eqref{eq.evs}  can be considerably mitigated in the limit  $T\rightarrow T_c$. In this limit, $\lambda(T)$ diverges, allowing one to approximate $f(k,z)\simeq \frac{1}{2}[e^{-kz}-e^{-k(z+d)}]$. Defining $I_\nu(a,b)=\int_0^\infty \!dx\,e^{-ax}J_\nu(bx)/x$ and using the identity $-\frac{\partial I}{\partial a}=\int_0^\infty \!dx\,e^{-ax}J_\nu(bx) = b^{-\nu}(\sqrt{a^2+b^2}-a)^\nu/\sqrt{a^2+b^2}$~\cite{gradshteyn2007}, we get
\begin{subequations}\label{eq.BzHighT}
  \begin{align}
    \label{eq.BzHighTa}
    B_z(r,z) \simeq & \frac{\phi_0}{4\pi\lambda^2}\int_0^\infty dk\,J_0(kr)\frac{e^{-kz}-e^{-k(z+d)}}{k} = \frac{\phi_0}{8\pi\lambda^2} \ln
    \left[
    \frac{\sqrt{r^2+(z+d_\text{S})^2}+z+d_\text{S}}{\sqrt{r^2+(z+d_\text{S})^2} - z-d_\text{S}}\cdot
    \frac{\sqrt{r^2+z^2}-z}{\sqrt{r^2+z^2} + z}
    \right],\\
    \label{eq.BzHighTb}
    B_r(r,z) \simeq & \frac{\phi_0}{4\pi\lambda^2}\int_0^\infty dk\,J_1(kr)\frac{e^{-kz}-e^{-k(z+d)}}{k} = \frac{\phi_0}{4\pi\lambda^2}
    \frac{\sqrt{r^2+z^2} + d - \sqrt{r^2+(z+d_\text{S})^2}}{r}.
  \end{align}
\end{subequations}
These equations allows for a much simpler, yet accurate, numerical estimate of $\epsilon_\text{vs}$ in the limit $T\rightarrow T_c$. For an analytical estimate, further simplification is necessary. Here, we estimate $\epsilon_\text{vs}$ in the limit of small skyrmion sizes, in which case the components of the magnetic induction for $r<R_\text{sk}$ can be approximated by:
\begin{equation}\label{eq.BzHiTr=0}
    B_z(r,z) = \frac{\phi_0}{4\pi\lambda^2} \ln\frac{z+d_\text{S}}{z} + {\cal O}(r^2)
    \quad \text{and} \quad
    B_r(r,z) = \frac{\phi_0}{8\pi\lambda^2} \left(\frac{r}{z}-\frac{r}{z+d} \right) + {\cal O}(r^3)
\end{equation}
\begin{figure}[b]
\centering
\includegraphics[width=\linewidth]{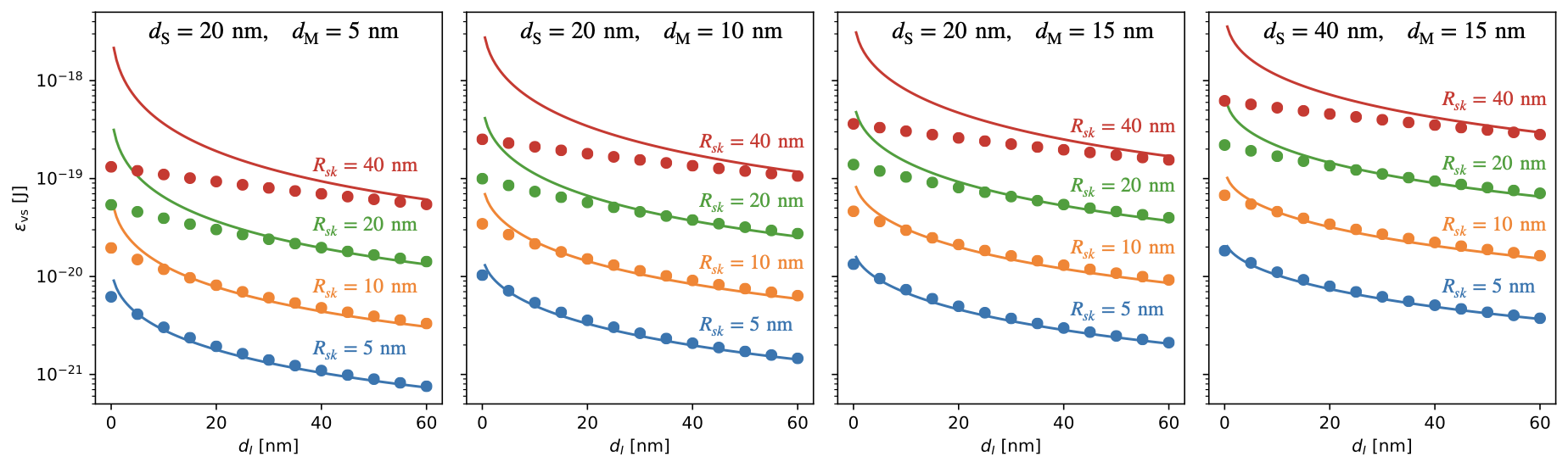}
\caption{High-temperature limit of the skyrmion-vortex coupling energy, $\epsilon_\text{vs}$, as a function of $d_\text{I}$ for different values of the skyrmions radius $R_\text{sk}$, $d_\text{S}$, and $d_\text{M}$. Symbols are numerically calculated using Eqs.~\ref{eq.evs}, \ref{eq.theta}, \ref{eq.BzHighTa}, and \ref{eq.BzHighTb}, and lines correspond to the analytical estimate Eq.~\ref{eq.evs-estimate}. We used $\xi_s=R_\text{sk}$, and $M_s=1.0$ MA/m, $\lambda=150$ nm, and $T=0.95T_c$. }
\label{fig.evsAnalytical}
\end{figure}
Additionally, we simplify the skyrmion profile as a linear function: $\theta=\pi(1-r/2R_\text{sk}) H(2R_\text{sk}-r)$, where $H(x)$ is the Heaviside step function. In this case, from~\eqref{eq.evs},  $\epsilon_\text{vs}=-2\pi  M_s\int_{d_\text{I}}^{d_\text{I}+d_\text{M}}[\int_0^{2R_\text{sk}}(B_z\cos\theta - B_r\sin\theta -B_z)\,rdr]dz$. Substituting the approximate form of $B_z$ and $B_r$ and performing the integrations, we get
\begin{equation}\label{eq.evs-estimate}
\epsilon_\text{vs} \simeq \frac{\phi_0M_sR_\text{sk}^2}{\lambda^2}
[d_\text{S}{\cal R}_1 (d_\text{S},d_\text{I},d_\text{M}) + R_\text{sk}{\cal R}_2(d_\text{S},d_\text{I},d_\text{M})].
\end{equation}
where ${\cal R}_1(d_\text{S},d_\text{I},d_\text{M})$ and ${\cal R}_2(d_\text{S},d_\text{I},d_\text{M})$ are  dimensionless constants depending only on the geometrical parameters of the superconductor-insulator-chiral magnet setup:
\begin{align}\label{eq.calR1}
  {\cal R}_1(d_\text{S},d_\text{I},d_\text{M}) =&~ c_1\left[
  \ln\left(\frac{d_\text{I}+d_\text{M}}{d_\text{I}}\right) + \frac{d_\text{S}+d_\text{I}+d_\text{M}}{d_\text{S}}\ln\left(\frac{d_\text{S}+d_\text{I}+d_\text{M}}{d_\text{I}+d_\text{M}}\right) - 
  \frac{d_\text{S}+d_\text{I}}{d_\text{S}}\ln\left(\frac{d_\text{S}+d_\text{I}}{d_\text{I}}\right)
  \right], \\
  {\cal R}_2(d_\text{S},d_\text{I},d_\text{M}) =&~ c_2\left[
  \ln\left(\frac{d_\text{I}+d_\text{M}}{d_\text{I}}\right) - \ln\left(\frac{d_\text{S}+d_\text{I}+d_\text{M}}{d_\text{S}+d_\text{I}}\right) \right].
\end{align}
with $c_1=2-8/\pi^2=1.189$ and $c_2=4(\pi^2-4)/\pi^3=0.757$. As shown in Fig.~\ref{fig.evsAnalytical}, this estimate provides an excellent approximation to $\epsilon_\text{vs}$ for small skyrmion radius. For larger $R_\text{sk}$, the approximation is still reasonable if $d_\text{I}\gtrsim R_\text{sk}$. 

Notice that this calculation assumes the vortex is aligned parallel with the polarization of the background ferromagnetic state, i.e., antiparallel to the skyrmion. This gives positive $\epsilon_\text{vs}$ and thereby repulsive skyrmion-vortex interaction. For the case where the vortex is aligned with the spin at the center of the skyrmion, one has to multiply Eqs.~\eqref{eq0a}-\eqref{eq.evs-estimate} by $-1$, which leads to negative $\epsilon_\text{vs}$, i.e. attractive skyrmion-vortex interaction.

Substituting Eq.~\eqref{eq.evs-estimate} in $\gamma^2=\epsilon_\text{vs}^2/(\epsilon_\text{vv}\epsilon_\text{ss})$ and using $\epsilon_\text{vv}=\phi_0^2d_\text{S}/(2\pi\mu_0\lambda^2)$ and $\epsilon_\text{ss}=60d_\text{M}R_\text{sk}^4/\xi_\text{s}^4$, one gets the estimate on the right-hand side of Eq. (9). Notice that, although Eq.~\eqref{eq.evs-estimate} provides an excellent approximation for small skyrmions, the accuracy of Eq. (9) is limited by the accuracy of the formula for $\epsilon_\text{ss}$. Since this formula is only order-of-magnitude accurate~\cite{Capic2020}, our estimate for $\gamma^2$ is to be taken as an order-of-magnitude approximation, even for small skyrmions. However, it can be promptly adjusted as soon as a more accurate expression for $\epsilon_\text{ss}$ becomes available.

 
\section{Additional supporting data}

\subsection{Phase diagram as a function of skyrmion density}
\label{sec.density}

\begin{figure*}[tbh]
\centering
\includegraphics[width=\linewidth]{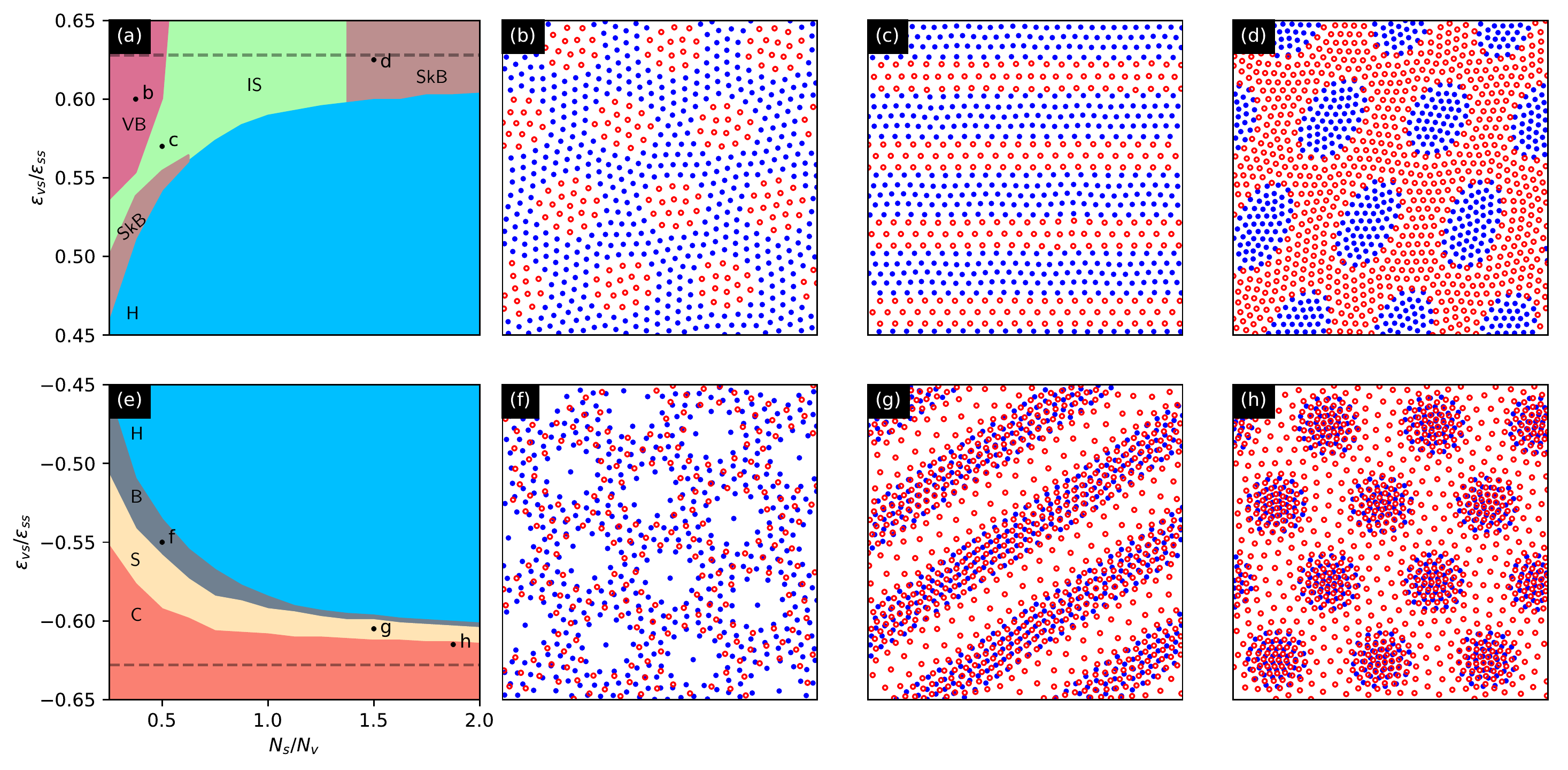}
\caption{(a) and (e) Phase diagram of skyrmion-vortex matter in the plane defined by the skyrmion-vortex coupling $\epsilon_\text{vs}$ and skyrmion concentration related to the number of vortices, $N_\text{s}/N_\text{v}$, for $\epsilon_\text{vs}>0$ (a) and $\epsilon_\text{vs}<0$ (b) . Fixed parameters are: $\xi_\text{s}=0.5\lambda_\text{vs}$, $\epsilon_\text{vv}=0.25\epsilon_\text{ss}$, $N_\text{v}=512$, and $L=12\lambda_\text{vs}$. (b)-(d) Representative configurations of all identified inhomogeneous phases are shown in (b)-(d), for repulsive s-v coupling, and (f)-(h), for attractive s-v coupling. Phase labeling and color scheme are the same as in Fig. 2.}
\label{fig.PD2}
\end{figure*}

In superconducting thin films, one typically has $\lambda>100$ nm, so that the condition $n_\text{v0}>1/\lambda^2$ is satisfied for fields as low as $\phi_0/\lambda^2=0.2$ mT. In contrast, skyrmions can be found at low densities $n_\text{s}\ll\xi_s$ under high or low fields, depending on the particular choice of chiral magnetic material and preparation conditions of the magnetic state. For such low skyrmion concentrations, the mean-field approximation is expected to fail. To check this and to determine the possible phases at low skyrmion concentrations, we performed a series of simulations for skyrmion densities ranging from $n_\text{s0}=7.12\lambda_\text{vs}^{-2}=1.76\xi_\text{s}^{-2}$, down to $n_\text{s0}=0.89\lambda_\text{vs}^{-2}=0.22\xi_\text{s}^{-2}$. The results, are shown in Fig.~\ref{fig.PD2}. 
All 7 phases identified in the diagram of Fig.~2 are also observed when changing the skyrmion concentration. Remarkably, the value of  $|\epsilon_\text{vs}|$ corresponding to the boundary of the homogeneous phase decreases considerably at low skyrmion concentrations, suggesting that modulated vortex-skyrmion phases are easier to observe in this case. For high skyrmion concentration, this boundary saturates at a constant value of $|\epsilon_\text{vs}|$ close to the mean-field stability line.

\newpage
\subsection{Equilibrium configurations for larger simulation box}

\begin{figure*}[b]
\centering
\includegraphics[width=0.8\linewidth]{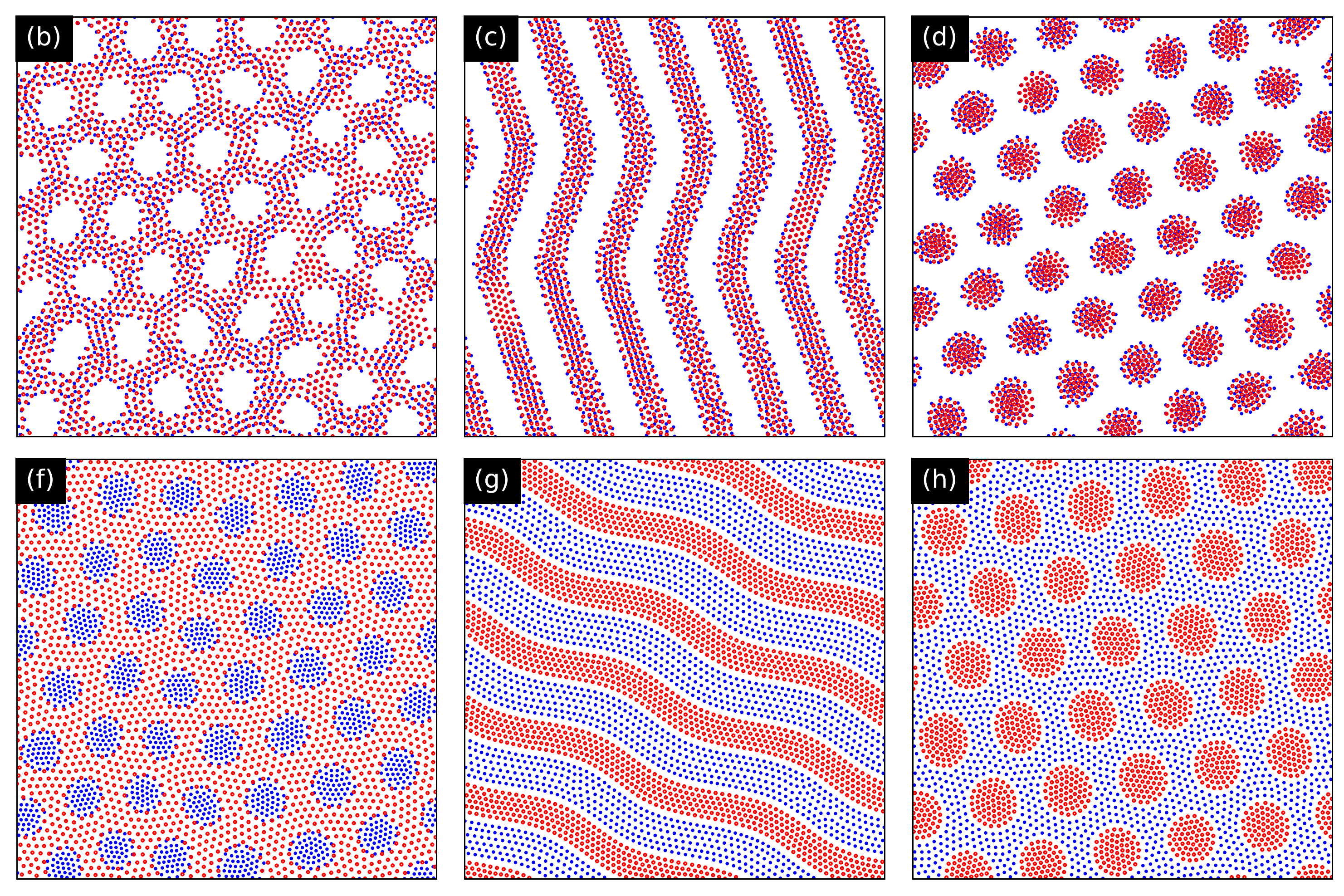}
\caption{Configurations obtained by simulated annealing for $N_\text{v}=2048$ and $N_\text{s}=2176$ in a simulation box of size $L=24\lambda_\text{vs}$ at the points of the phase diagram indicated in Fig.~2-(a). The parameters are the same as for Figs.~2-(b)-(d) and (f)-(h), except for the size of the simulation box. 
}
\label{fig.Confs_L=24}
\end{figure*}

To identify possible size effects, we repeated the simulated annealing procedure at some points of the diagram of Fig. 2-(a) for a larger system: $L=12\lambda_{vs}$, $N_v=2048$, and $N_s=2176$. We found no noticeable change in the phase boundaries for the different system sizes. However, as expected, the distortion of the triangular cluster lattices induced by the periodic boundary conditions are considerably less pronounced for the larger system. In particular, we repeated the same annealing procedure at the points (b)-(d) and (f)-(h) of indicated in Fig.~2-(a), but for the larger system size. The results are shown in Fig.~\ref{fig.Confs_L=24}.

The enlarged simulation box is crucial for a more accurate determination of the wavelength $\ell$ of the mesoscale patterns, shown in Fig. 3. The $L=12$ simulation box imposes a small discrete set of values of $\ell$, which can be understood as follows. The orientation of straight stripes in a square simulation box of side $L$ can be determined by a simple trigonometry analysis, which gives $\sin \theta = \ell / (L/n_x)$ and $\cos \theta = \ell / (L/n_y)$, where $n_x$ ($n_y$) is the number of stripes cutting the $x$ ($y$) axis and $\theta$ is the stripe inclination angle with respect to $x$. Eliminating $\theta$ one has $\ell = L/\sqrt{n_x^2+n_y^2}$. The periodic boundary conditions force both $n_x$ and $n_y$ to be integers for any $L$, thereby imposing artificially a discrete set of values for $\ell$. For example, for $L=12\lambda_{vs}$ and considering points of the stripe phase close to the instability line (at which $2.7 \lesssim \ell \lesssim4.7$ as predicted by the mean-field theory),  only 7 values of $\ell$ are allowed, corresponding to the $(n_x,n_y)$ pairs: (0,3), (0,4), (1,3), (1,4), (2,2), (2,3), (3,3). Permutation of indices in a  $(n_x,n_y)$ pair results in the same value of $\ell$. The inclination angles in Figs.~2-(c), 2-(g), S2-(c), and S2-(g) are, respectively, 56.3\deg, 45\deg, 0.0\deg, and 33.7\deg, which correspond to $(n_x,n_y)= (3,2)$, (2,2), (0,4), and (2,3).  Notice that 2-(c) and S2-(g) have the same value of $\ell$.

When we double the system size ($L=24$), the number of possible values of $\ell$ in the range $2.7 < \ell < 4.7$ increases to 22. In addition, the larger system size also allows for undulated stripe patterns, which are not bounded by discrete values determined by the pairs $(n_x,n_y)$. This makes the $L=24$ system much less influenced by the boundary conditions.

\subsection{Relaxational dynamics and metastable states}

The phase diagrams presented in Figs.~(2) and \ref{fig.PD2} were constructed by minimizing the system free-energy following a simulated annealing procedure. The final minimum-energy configuration does not depend on the details of the dynamics. However, typical experiments are conducted at low temperatures, where fluctuations are negligible. Moreover, when decreasing the temperature from above the superconducting critical temperature $T_c$ under a fixed applied field, vortices nucleate just below $T_c$ in a situation where the skyrmions are already arranged in an approximately triangular lattice. In this case, the final state resulting from the interaction between the newly born vortices and the skyrmions can be metastable and strongly dependent on dynamical details like the friction coefficients, $\eta_\text{v}$ and $\eta_\text{s}$, and the gyromagnetic constant, $G$, appearing in Eqs.~(7) and (8). In particular, $G/\eta_\text{s}$ is a measure of the influence of the Magnus force on the skyrmion dynamics. To investigate this, we solved Eqs.~(7) and (8) for $N_\text{v}=2048$ and $N_\text{s}=2176$ in a simulation box of size $L=24\lambda_\text{vs}$, fixing $G=4\pi\eta_v$ and four different values of $\eta_s$. We also fixed $\xi_\text{s}=0.5\lambda_\text{vs}$, $\epsilon_\text{vs}=1.2\epsilon_\text{ss}$, and $\epsilon_\text{vv}=0.7\epsilon_\text{ss}$. For these parameters, the system is expected to be in the intercalated stripe phase in equilibrium [see Fig.~(2)]. 

\begin{figure*}[b]
\centering
\includegraphics[width=\linewidth]{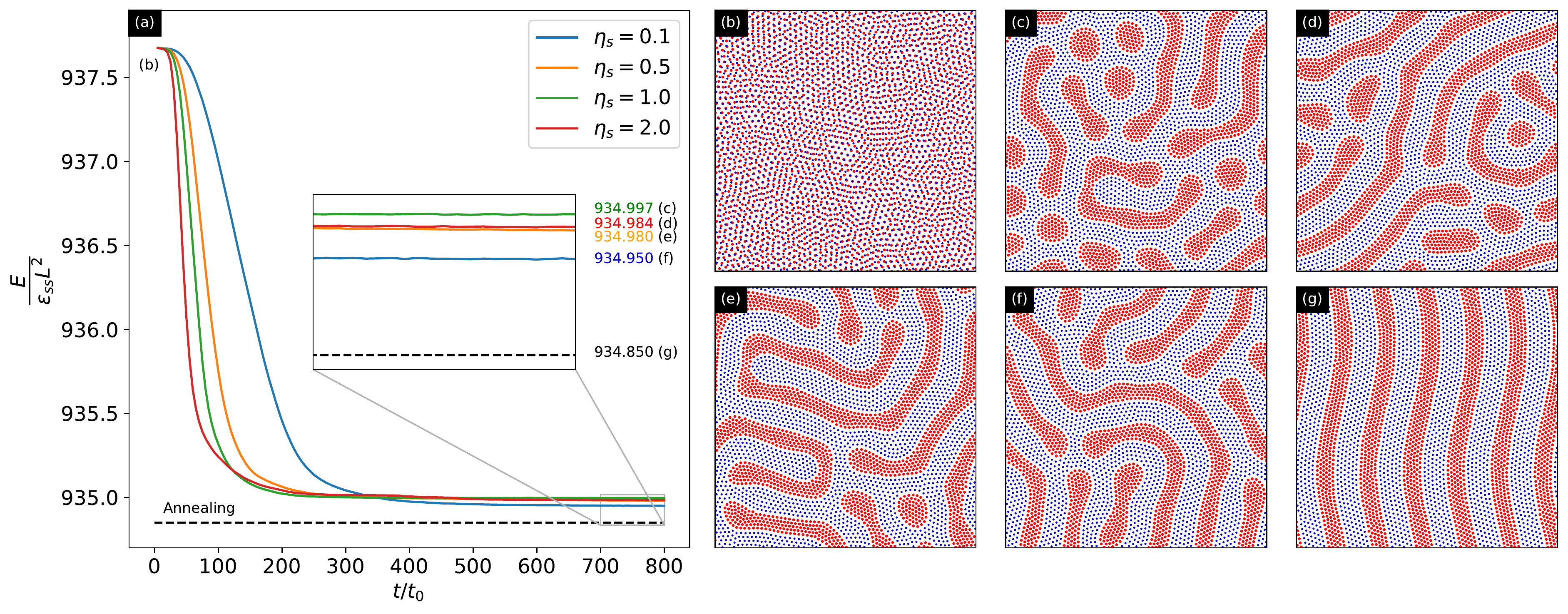}
\caption{(a) Time dependence of the energy per unit area of a vortex-skyrmion system ($N_\text{v}=2048$, $N_\text{s}=2176$, and $L=24\lambda_\text{vs}$) during relaxation from a homogeneous state (b) toward mesoscale, phase-separated states (c-f). Parameters were chosen so that the equilibrium configuration of the system is in the intercalated-stripe phase (g). The energy values of the final ($t=800t_0$) states are indicated in the zoomed in region (inset) of panel (a) and compared against the energy of the equilibrium (annealed) configuration (dashed horizontal line). 
}
\label{fig.RD}
\end{figure*}

To simulate a quench from above the superconducting temperature toward the target situation represented by the chosen parameters, we initialize the skyrmions as a triangular lattice and the vortices at random positions and record the subsequent time evolution of the system with the Gaussian noises turned off. The results are presented in Fig.~\ref{fig.RD}.
At early times ($t\sim t_0=\eta_v\lambda_\text{vs}^2/\epsilon_\text{ss}$), vortices rapidly arrange in an approximately triangular lattice for all $\eta_s$ investigated, so that the vortex-skyrmion system is momentarily in the homogeneous state (b). As time evolves further, a phase separation process starts and the homogeneous state decays to different inhomogeneous configurations (c-f), where the system finally settles with an energy considerably lower than that of the homogeneous phase. All these configurations are metastable: their energy are slightly higher than that of the equilibrium configuration obtained by simulated annealing (g), which corresponds to the stripe phase. Notice that the decay time is larger for smaller skyrmion friction coefficients. This counterintuitive result can be explained as follows: for low $\eta_\text{s}$, the Magnus force dominates the dynamics, forcing the skyrmions to perform spiral rather than straight trajectories toward local minima~\cite{RaiJose2019}. This increases the time for the skyrmions to reach local equilibrium, retarding the relaxation of the whole system. Interestingly, the jiggling produced by the spiralling dynamics seems to produce an annealing effect, as the final configurations for smaller $\eta_\text{v}$ have in general lower energy as compared to the other metastable states. 

\section{Validity and limitations of the point particle model}


Vortices can be treated as pointlike particles within the so-called London approximation, valid when intervortex distances are large compared to the coherence length $\xi$, which is a measure of the vortex core radius. A similar approximation also applies to skyrmions, when their mutual separation is large compared to the skyrmion radius $R_{\rm sk}$. However, in contrast to vortices, skyrmions are highly sensible to the local flux density. 
They can shrink and eventually colapse when exposed to a high flux density. Conversely, at low fields, they can increase in size and eventually merge into spirals. Therefore, it is important to check the possibility that the phase separation process predicted in our work does not conserve the total number of skyrmions and/or is accompanied by drastic changes in the skyrmion morphology, thus invalidating our approach.



\begin{figure*}[ht]
\centering
\includegraphics[width=\linewidth]{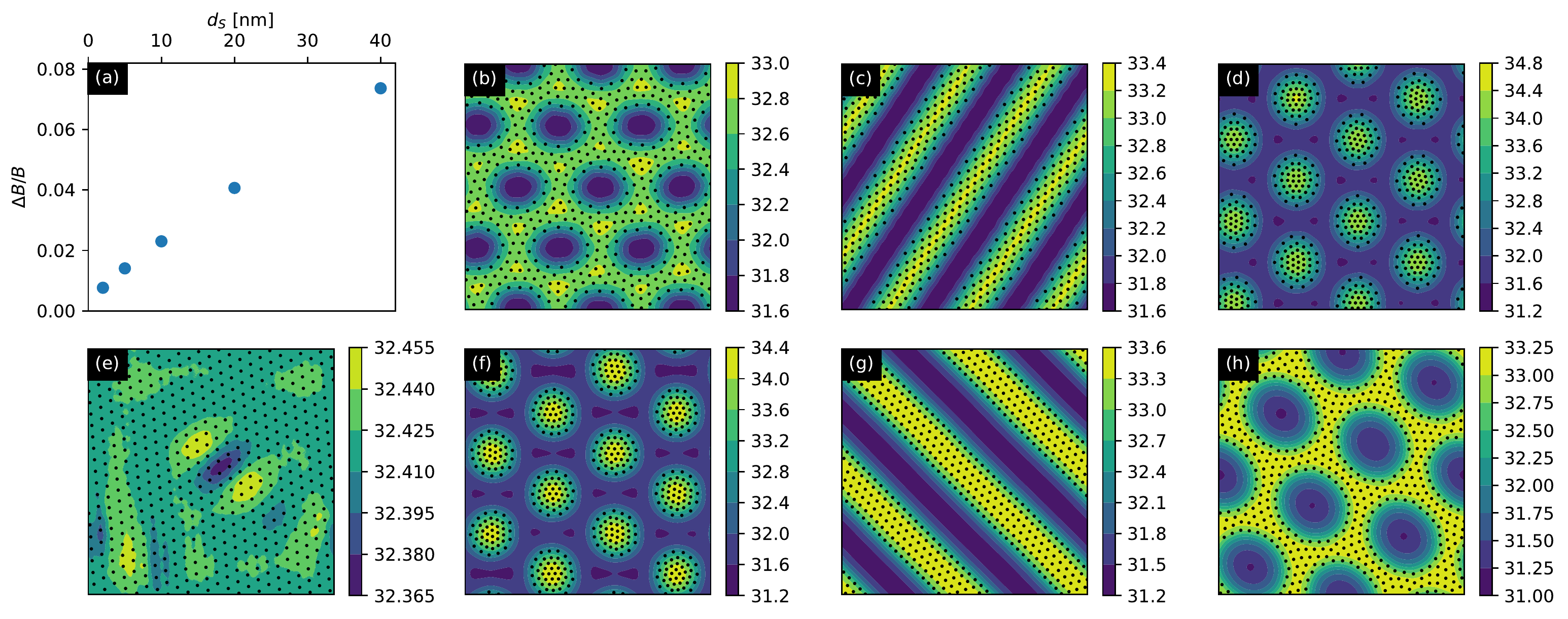}
\caption{(a) Relative amplitude of the modulation of the magnetic flux texture induced by the vortex configuration of Fig. 2-(g) calculated at the surface of the superconducting film assuming different values of the film thickness $d_{\rm S}$ ranging from 2 to 40 nm. (b)-(h) Contour plots of the magnetic field profiles [in units of $\phi_0/(2\pi\lambda^2)$] induced by the vortex configurations shown in Fig. 2 (b)-(g) (here represented by the black dots) at a height 15 nm from the SC film surface. Here we assumed $d_{\rm S}=20$ nm.}
\label{fig.FieldProfile}
\end{figure*}

In Fig.~\ref{fig.FieldProfile}, we present the field profile $B_z(x,y,z_0)$ induced by the vortex distributions of Fig. 2. The profiles were calculated at a height $z_0=15$ nm from the SC film surface using Eq.~S3 and assuming $d_{\rm S}=20$ nm. They reveal that the flux density only ripples about the mean value [$\langle B_z\rangle=32.4 \phi_0/(2\pi\lambda^2)$] with a relative amplitude typically smaller than 5\%. For a height $z_0=0$, the maximum amplitude raises to 7\%. This smoothness of the flux distribution induced by the  modulated vortex patterns is a result of the long range nature of the stray fields emerging from the vortices and the small (mesoscopic) length scale of the density modulations. It can be less than 1\% for ultrathin SC films as shown in Fig.~\ref{fig.FieldProfile}-(a). In this case, if one removes the insulating layer, the spin orbit coupling between the SC and CM layer may dominate the skyrmion-vortex interaction, opening an interesting perspective for the realization of attractive skyrmion-vortex interaction even with the external field aligned with the background magnetization of the chiral magnet. Such a situation would be a many vortex--many skyrmion extension of the skyrmion-vortex structure studied in Ref. [20] and could be interesting for quantum computing applications.

To address the question whether magnetic field textures similar to those obtained in our simulations can induce significant change in the morphology of the skyrmions, we performed micromagnetic simulations of a chiral magnetic film subjected to a static nonuniform magnetic field, modeled as $B_z(x)= B + \Delta B \sin(\frac{2\pi x}{\ell})$, thus imitating the field profile induced by the mesoscale vortex distribution. The simulations performed for homogeneous field $\bm{B}=B\hat{\bm{z}}$ revealed  the usual three phases: the spiral phase, for $B<0.2$ T, skyrmion crystal, for $0.2$ T $<B<0.7$ T, and the fully polarized ferromagnetic state for $B>0.7$ T. 
\begin{figure*}[ht]
\centering
\includegraphics[width=\linewidth]{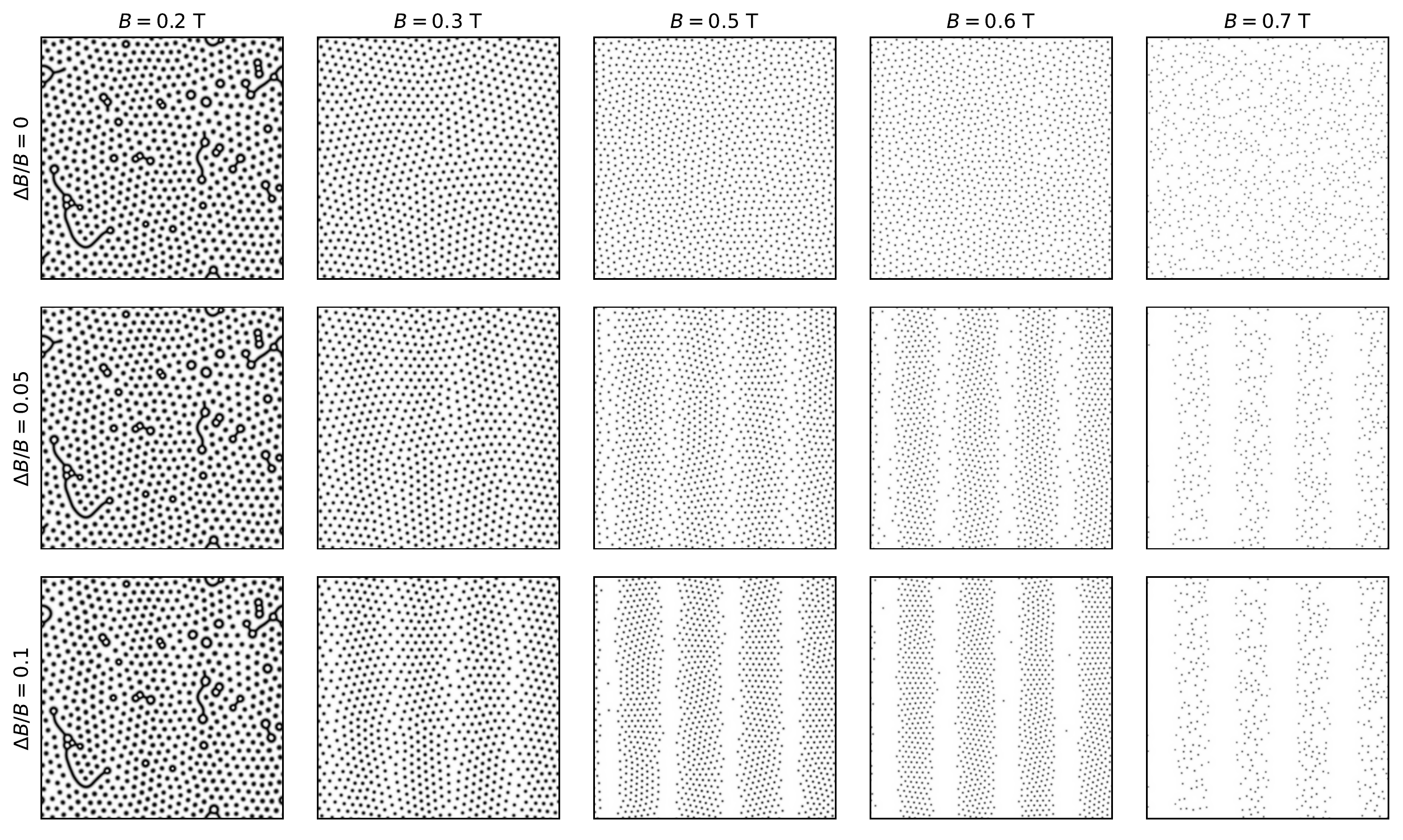}
\caption{Contour plots of the off plane magnetization ($m_z$) of a chiral  magnetic film subjected to uniform (top row) and nonuniform (middle and bottom rows) magnetic fields.  The micromagnetics simulations were performed on a square box of lateral size $L=1.44$ $\mu$m with periodic boundary conditions using the package MuMax3~\cite{Vansteenkiste2014TheMuMax3}. The chosen period of the field modulation was  $\ell=360$ nm. The magnetic parameters are: saturation magnetization $M_s=1.0$ MA/m, exchange stiffness $A=10.45$ pJ/m, DMI $D=3$ mJ/m$^2$, and effective anisotropy $K_{\rm eff}=0.09$ MJ/m$^3$.}
\label{fig.micromagnetics}
\end{figure*}

\begin{figure*}[ht]
\centering
\includegraphics[width=0.355\linewidth]{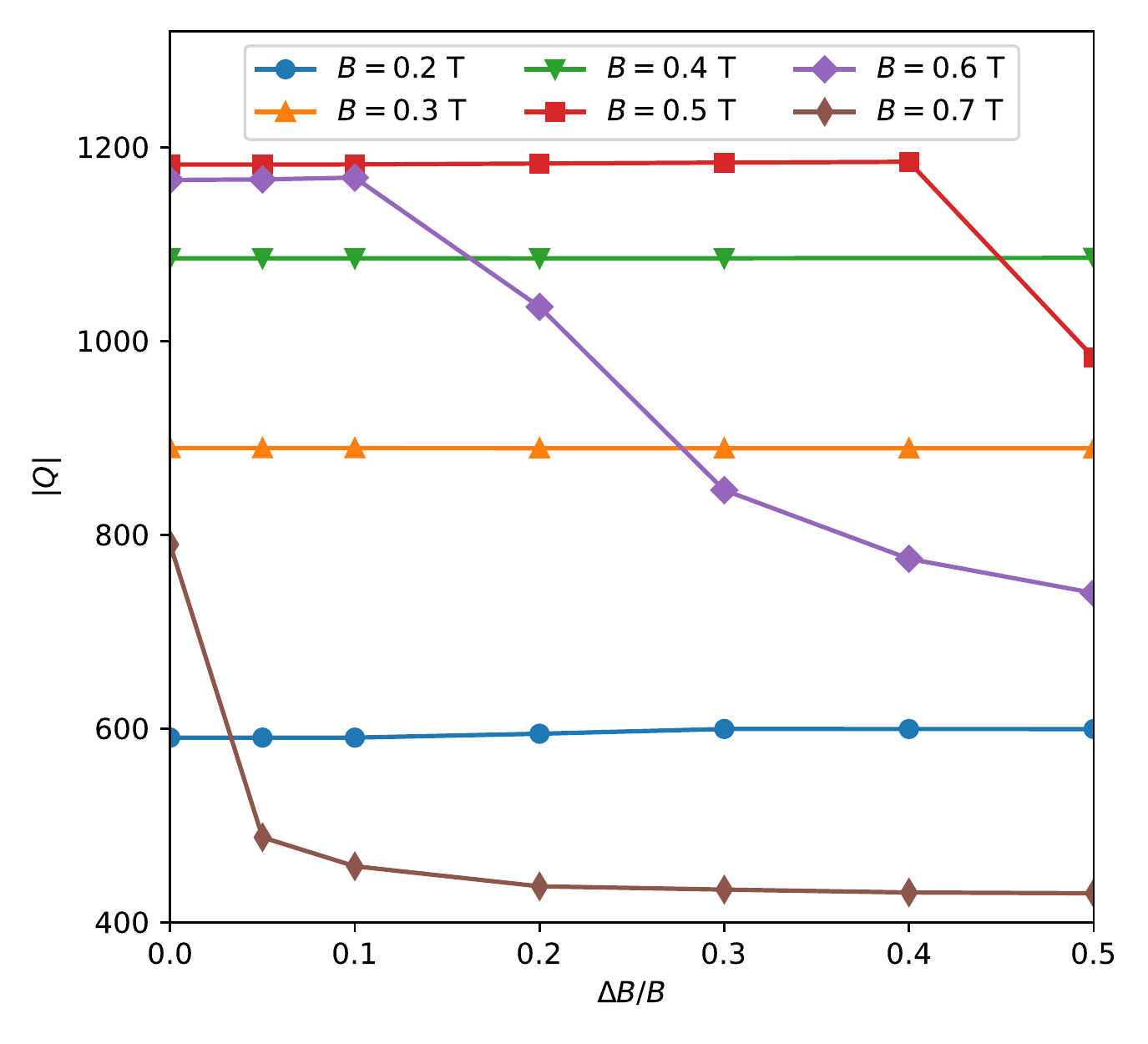}
\caption{Total topological charge of the magnetic film of Fig.~\ref{fig.micromagnetics} as a function of the relative amplitude of the spatial modulations of the external magnetic field. In our molecular dynamics simulations, the typical values of $\Delta B/B$ is less than 0.1.}
\label{fig.Q}
\end{figure*}

The configurations shown in Fig.~\ref{fig.micromagnetics} for several values of $B$ and $\Delta B$ reveal that, in the skyrmion crystal phase, the main effect of the added inhomogeneity is to displace skyrmions from regions of higher to lower local flux density, while preserving the total number of skyrmions, similar to our molecular dynamics simulations. However, for $B=0.7$ T (that is, at the vicinity of the ferromagnetic phase), where skyrmions are found in a sparse disordered distribution, the  inhomogeneity induces the collapse of the skyrmions at the regions of higher local magnetic field, while those at the regions of lower field remain frozen in, but with a somewhat larger size. 

We also calculated the total topological charge of the magnet, $Q=\frac{1}{4\pi}\int \hat{\bm{n}}\cdot\left(\partial_x\hat{\bm{n}}\times\partial_y\hat{\bm{n}}\right)dx\,dy$, as a function of the relative amplitude of the inhomogeneity for several values of $B$ (Fig.~\ref{fig.Q}). In the skyrmion phase, $|Q|$ is a measure of the number of skyrmions, $N_s$, in the system. For $0.2< B\leq0.6$, that is, in the full skyrmion crystal phase, $|Q|$ is essentially constant for small $\Delta B/B$, while for $B=0.7$ T a drastic change in the number of skyrmions was observed already for the smallest value of $\Delta B$. 
\newpage

These results attest the validity of our point particle approach for the case where the chiral magnet is within the skyrmion crystal phase. Outside this region, our model does not apply since, at low field, the system becomes unstable with respect to the formation of spirals (like the worms and loops seen in the first column of Fig.~\ref{fig.micromagnetics}) while at high fields ($B\geq0.7$ T) the number of isolated skyrmions is not conserved. Conservation of skyrmion number in the phase separation process is particularly crucial to our model, since our simulations were performed for constant $N_v$ and constant $N_s$. The conservation of vortex number, $N_v$, is guaranteed by flux quantization within the simulation area, which is assumed to be exposed to a constant macroscopic field.

\bibliography{references}